\documentclass[aps,twocolumn, floatfix,superscriptaddress]{revtex4-2}
\usepackage{graphicx}
\usepackage{hyperref}
\usepackage{amsmath}
\usepackage{amsfonts}
\usepackage{mathtools,bm}
\usepackage[amssymb]{SIunits}
\usepackage{color}
\usepackage{mathrsfs}

\begin{document}
\flushbottom
\title{Crossover Between Quantum and Classical Waves and High Frequency Localization Landscapes}

\author{David Colas}
\affiliation{Aix Marseille Univ, CNRS, Centrale Marseille, LMA UMR 7031, Marseille, France}

\author{Cédric Bellis }
\affiliation{Aix Marseille Univ, CNRS, Centrale Marseille, LMA UMR 7031, Marseille, France}

\author{Bruno Lombard}
\affiliation{Aix Marseille Univ, CNRS, Centrale Marseille, LMA UMR 7031, Marseille, France}

\author{Régis Cottereau}
\email{cottereau@lma.cnrs-mrs.fr}
\affiliation{Aix Marseille Univ, CNRS, Centrale Marseille, LMA UMR 7031, Marseille, France}

\begin{abstract}

Anderson localization is a universal interference phenomenon occurring when a wave evolves through a random medium and it has been observed in a great variety of physical systems, either quantum or classical. The recently developed localization landscape theory offers a computationally affordable way to obtain useful information on the localized modes, such as their location or size. Here we examine this theory in the context of classical waves exhibiting high frequency localization and for which the original localization landscape approach is no longer informative. Using a Webster's transformation, we convert a classical wave equation into a Schr\"{o}dinger equation with the same localization properties. We then compute an adapted localization landscape to retrieve information on the original classical system. This work offers an affordable way to access key information on high-frequency mode localization.

\end{abstract}

 \date{\today} \maketitle

Anderson localization refers to wave localization due to the presence of a strongly inhomogeneous medium. Originally predicted by P. W. Anderson in 1958 for electronic wave functions~\cite{anderson58a}, this phenomenon was since evidenced in a great variety of oscillatory systems, such as electromagnetic~\cite{laurent07a,schwartz07a,riboli10a} or matter waves ~\cite{piraud15a,billy08a,white20a}, in the context of meta-materials~\cite{asatryan10a}, photonic lattices~\cite{lahini08a} or cavity QED~\cite{sapienza10a}. 
Anderson localization has also been studied for classical vibrating systems, mostly for ultrasounds~\cite{condat87a,weaver90a,lobkis08a,hu09a,dhillon21a}.
Notable differences exist in the mathematical structure of the operators describing classical and quantum waves, and the general properties of Anderson localization, such as the link between the presence of spectral gaps and the emergence of localized modes, are still discussed~\cite{kirkpatrick85a,figotin96a,altmann20a}. Despite this abundant literature produced over six decades, many questions on the nature of Anderson localization remain open. One important concern is the following: is it possible to determine, from the knowledge of the random medium configuration, where waves are going to localize? And this, without solving the computationally expensive associated eigenvalue problem. 

In 2012, M. Filoche and S. Mayboroda developed an original and innovative tool to apprehend
wave localization, which they coined as the \textit{localization landscape}~\cite{filoche12a}.
For a given quantum potential $V(x)$, possibly random, one considers the eigenvalue problem for the
Schr\"{o}dinger equation in a domain $\Omega$
\begin{equation}
(-\Delta + V(x)) \psi_n(x)= E_n \psi_n(x)\,,\hspace{10mm} \psi_n\big|_{\partial\Omega}=0\,.
  \label{eq:1}
\end{equation}
Instead of Eq.(\ref{eq:1}), one can solve the elliptic equation
\begin{equation}
(-\Delta + V(x))u_\psi(x) = 1\,,\hspace{22mm} u_\psi\big|_{\partial\Omega}=0\,,
  \label{eq:2}
\end{equation}
with $u_\psi$ the so-called Localization Landscape (LL). 
When analyzing the LL's shape, \textit{i.e.} its peaks and valleys, one can predict the position, shape and energy of the localized eigenstates, to a certain extent. The LL can actually be understood as the inverse of an effective confining potential for the eigenmodes~\cite{arnold16a}, and it has been successfully applied in various contexts~\cite{filoche17a,piccardo17a,li17a,chalopin19a,balasubramanian20a,lemut20a,garcia21a,shamailov21a}.

This remarkable theory is also extendable to localization associated with other symmetric elliptic differential operators governing wave propagation, such as the Laplacian, the bilaplacian or the operator $-\textrm{div}(A(x)\nabla)$, notably describing classical waves in inhomogeneous media. However, as we will show, the LL only returns useful information in the case of low frequency localization. When the system exhibits high frequency localization, \textit{i.e.} when the first eigenmodes are delocalized, the LL does not bring any insight on the position or shape of the localized modes. Such situation might occur upon the choice of the differential operator or the boundary conditions. 

In this Letter, we adapt the Localization Landscape Theory (LLT) to classical scalar waves exhibiting high frequency localization, using the example of a classical wave equation such as the Helmholtz equation governing acoustics. For a given type of random quantum potential or acoustic structure, we first review and compare the phenomenology associated with the quantum and classical systems, which is essentially an opposite transition between the localization and delocalization regimes. Then, we make use of the Webster's transformation~\cite{webster19a,martin04a} to symmetrize and convert the Helmholtz equation into a Schr\"{o}dinger equation with an equivalent random potential that preserves the localized nature of high-energy modes. From that, we build an adapted LL to retrieve information on the original classical modes and we calculate their minimal support. This work highlights limitations of the original LLT for systems exhibiting high frequency localization and enriches it by restoring the LL as a predictive and computationally affordable tool to treat Anderson localization.

\begin{figure*}[t!]
  \includegraphics[width=\linewidth]{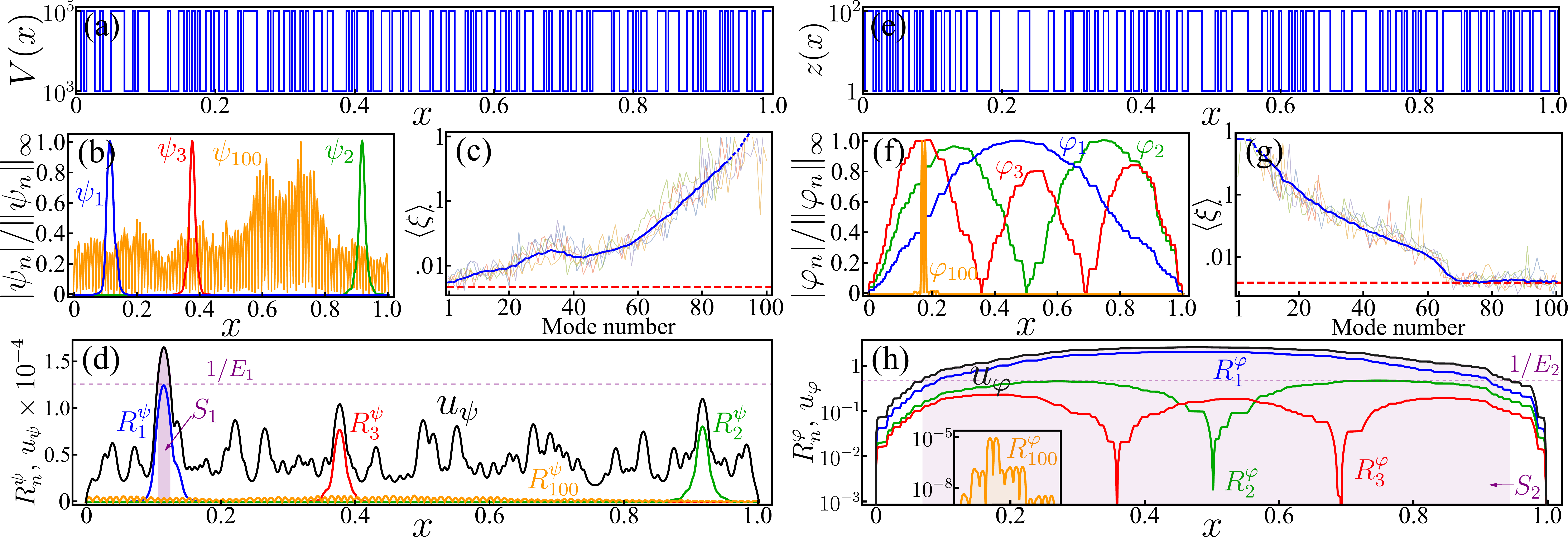}
  \caption{Comparison of mode localization between a quantum (a-d) and a classical scalar system (e-h). (a,e) Random Bernoulli-type quantum potential $V(x)$ and classical parameter $z(x)$ with 256 sites. (b,f) Selection of eigenmodes normalized to their max, for the quantum and classical system, respectively. (c,g) Mode localization length $\langle \xi \rangle$, averaged on $1000$ realizations (blue lines) along with some single trajectories. The dash part of the blue curves indicates that the exponential fit is less relevant when modes are delocalized~\cite{foot1}. The dashed-red line set at $1/256$ defines the smallest possible localization site, hence the limit value for $\langle\xi\rangle$. (d,h) Same selection of modes $R^\psi_{n}$ and $R^\varphi_{n}$, now normalized to their energy, see Eq.(\ref{eq:3}), and bounded by the corresponding LLs $u_{\psi}$ and $u_{\varphi}$ (black lines). The purple-shaded areas correspond to the supports $S^\psi_{1}$ and $S^\varphi_{2}$ for the modes $\psi_1$ and $\varphi_2$, defined by Eq.(\ref{eq:8778}). An animation of panels (d,h) is provided in the Supplemental Material~\cite{SuppMat}.}
  \label{fig:1}
\end{figure*}
We start our analysis by examining the phenomenological differences between localization of quantum and classical waves. We first consider the 1D Schr\"{o}dinger eigenvalue problem of Eq.(\ref{eq:1}) with a Bernoulli-type random potential $V(x)$ constructed as follows: a unit-length space is divided into $N$ sites on which $V$ either takes the value $V_\mathrm{min}$ or $V_\mathrm{max}$, with equiprobability. An example of possible realization with 256 sites is shown in Fig.\ref{fig:1}(a). We then calculate the first $100$ eigenmodes and show a selection of them in Fig.\ref{fig:1}(b). Localized modes are expected to decay exponentially as $|\psi_n(x)|\sim \exp(-x/\xi)$, with $\xi$ the localization length. In Fig.\ref{fig:1}(c) we show the measurement $\langle\xi\rangle$ for each mode and averaged over a thousand realizations~\cite{foot1}. First modes are indeed localized and delocalization progressively occurs at higher energies. We then compute the LL $u_\psi$ from Eq.(\ref{eq:2}) and present it in Fig.\ref{fig:1}(d). One can see the effectiveness of the LLT, with the matching between the LL's main peaks and the first localized modes. The mathematical essence of the LLT is encapsulated in the following inequality~\cite{filoche12a}:
\begin{equation}
R^\psi_{n} \overset{\text{def}}{=} \frac{|\psi_n(x)|}{||\psi_n(x)||_{L^\infty} E_n}\leq u_\psi(x)\,,
\label{eq:3}
\end{equation}
which states that the LL $u_\psi>0$ acts as a supremum and bounds the eigenmodes, when normalised by their energy. In Fig.\ref{fig:1}(d), we show how the LL acts as an upper bound for the normalized modes $R^\psi_{n}$. Moreover, an eigenmode $\psi_n$ with energy $E_n$ possesses the following approximate support~\cite{arnold19a}: 
\begin{equation}
S^\psi_{n}\overset{\text{def}}{=}\left\{ x \in \Omega : u_\psi(x) \geq \frac{1}{E_n} \right\}\,.
\label{eq:8778}
\end{equation}
In other words, the support $S^\psi_{n}$ defines where most of a mode's ``mass'' must sit, as illustrated in Fig.~\ref{fig:1}(d). As $E_n$ increases with $n$, eigenmodes are allowed to occupy a larger region of space and thus to delocalize.  

We now perform the same analysis with a classical system, taking the example of the 1D scalar wave equation 
\begin{equation}
-\frac{1}{\rho(x)}\nabla\cdot(\kappa(x)\nabla \varphi_n(x))= \omega_n^2  \varphi_n(x)\,,
\label{eq:4}
\end{equation}
where $\rho(x)$ and $\kappa(x)$ can be interpreted as a mass density and bulk modulus in the context of acoustics, and $\omega_n$ the eigenfrequencies playing the role of an energy. Note that in this form, the operator in the LHS of Eq.(\ref{eq:4}) is not self-adjoint but could be rewritten as a generalized eigenvalue problem. We further simplify this problem by considering a single random variable, setting $\rho(x)=\kappa(x)=z(x)$. Such approximation is known as the Webster's horn equation~\cite{webster19a,martin04a} and as we will see later on, can be easily mapped into a Schr\"{o}dinger equation. The random parameter $z(x)$ is constructed in the same way as for $V(x)$, and can take the values $z_\mathrm{min}$ and $z_\mathrm{max}$, see Fig.\ref{fig:1}(e). A selection of eigenmodes for Eq.(\ref{eq:4}) is shown in Fig.\ref{fig:1}(f) along with the averaged measurement of their localization length $\langle\xi\rangle$. Unlike the Schr\"{o}dinger system, first modes are fully delocalized and localization progressively occurs at higher frequencies, see Fig.\ref{fig:1}(g), until $\langle \xi \rangle$ reaches the minimal possible value allowed by the system, that is the size of a single site $1/N$. In the Supplementary Material, we show that for a homogeneous 2D problem, the choice of boundary conditions can also lead to a modification of the localization regime~\cite{SuppMat}. 
 
The phenomenological difference between the quantum and classical systems can be intuitively understood from the structure of Eqs.(\ref{eq:1}) and (\ref{eq:4}). We refer to the argument presented in Ref.~\cite{vanderbaan01a}, which essentially states that for the Schr\"{o}dinger equation, the potential term is additive to the energy, so that a quantum wave predominantly perceives the effect of the disordered potential at low energies, thus exhibiting low energy mode localization. On the other hand, the classical wave equation can be rewritten in the form of a Laplacian plus another term, equivalent to a potential, but multiplicative to the energy. At low frequencies, a classical wave thus essentially perceives a homogeneous medium, and the first eigenmodes resemble the usual global modes of the domain~\cite{fouque_book07a}, see Fig.\ref{fig:1}(f). The random fluctuations become predominant only at higher frequencies, allowing Anderson localization.

We then attempt to compute an equivalent LL for the classical system, for this purpose we solve
%
\begin{equation}
-\frac{1}{z(x)}\nabla\cdot(z(x)\nabla u_\varphi(x))=1\,.
  \label{eq:6}
\end{equation}
and show the solution $u_\varphi$ in Fig.\ref{fig:1}(h) along with a selection of normalized modes $R^\varphi_{n}$, as previously done in Eq.(\ref{eq:3}). Despite the supremum $R^\varphi_{n}\leq u_\varphi$ being seemingly valid, $u_\varphi$ does not here provide any useful information about the mode localization. It is akin to the first eigenmode of the system, and does not possess any peaks or valleys. The higher-energy modes, regardless of their localization, rapidly fall orders of magnitude below $u_\varphi$. Equivalently, the support $S^\varphi_{n}$ for the first modes already represents a large region of space and quickly converges to the whole domain $\Omega$ as the eigenvalue increases, and so as $1/E_n$ decreases. The solution $u_\varphi$ thus fails to provide information when the first modes of the system are delocalized. 

\begin{figure}[t!]
  \includegraphics[width=\linewidth]{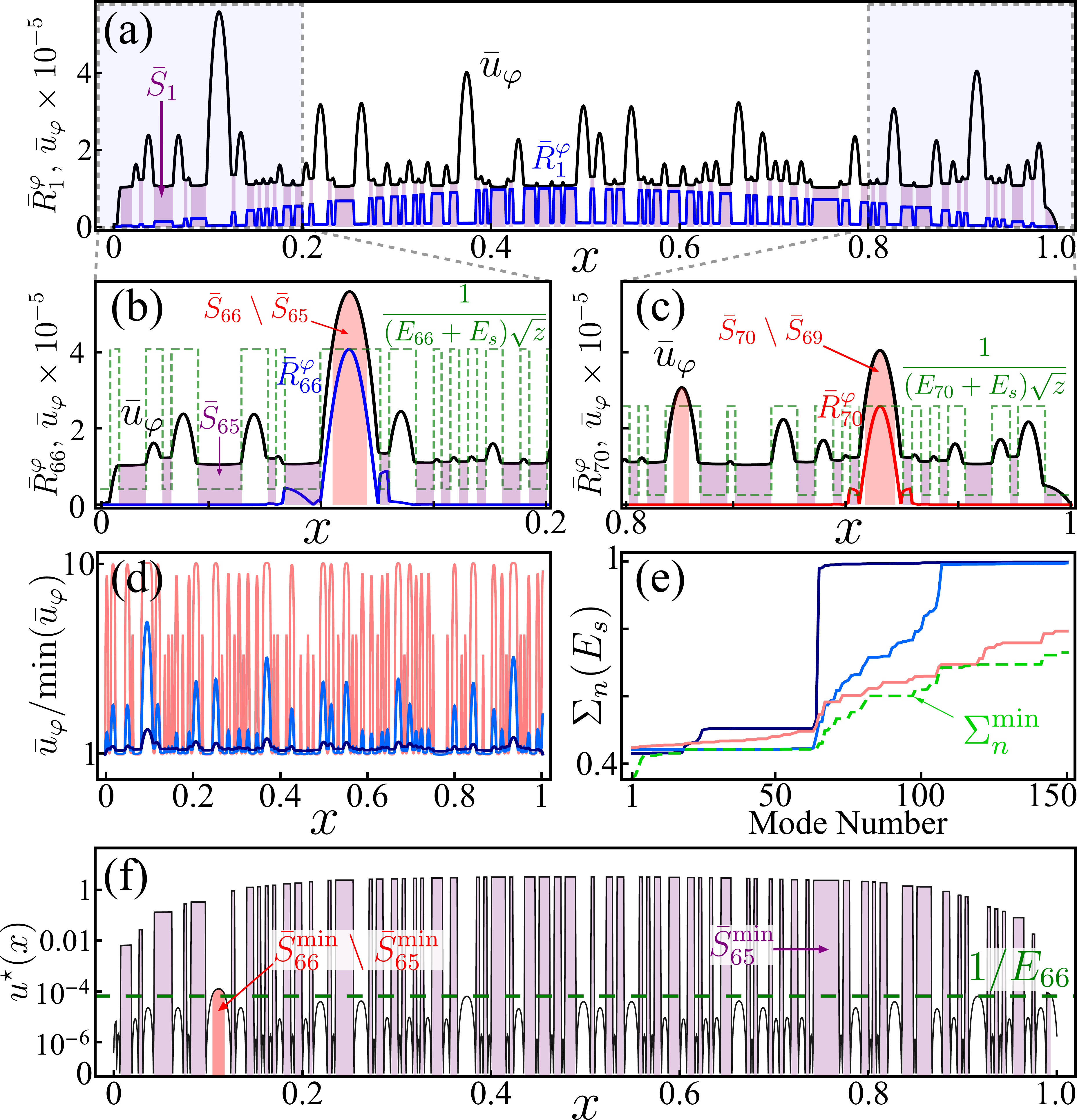}
  \caption{(a) Adapted LL $\bar{u}_\varphi(x;E_s)$ (black) computed with $E_s=2\times10^4$, and first eigenmode $\bar{R}^\varphi_1$ (blue). Purple-shaded areas correspond to the support $\bar{S}_1^\varphi$ defined in Eq.(\ref{eq:8787844}). (b,c) Focus on localized modes $\bar{R}_{66}^\varphi$ and  $\bar{R}_{70}^\varphi$. The dashed-green line corresponds to the RHS of the inequality in Eq.(\ref{eq:87878}) which locally defines the mode's support. The red-shaded area indicates the extra support obtained from the previous eigenvalue.
(d) Evolution of the adapted LL $\bar{u}_\varphi$ (re-scaled) for different values of energy shift $E_s \in \{10^3, 2\times10^4, 10^6 \}$ (blue to red).  (e) Measure of mode support $\Sigma_n^\varphi(E_s)$ for the same values of shifts $E_s$ as in (c), and measure of the minimum support $\Sigma_n^{\,\mathrm{min}}$ (dashed-green). (f) Function $u^\star(x)$ defining the minimal mode support (black). The purple-shaded area is the support for the $65^\mathrm{th}$ eigenvalue while the red-dashed area indicates the extra support obtained for the $66^\mathrm{th}$ one. An animation of panels (a,f) for a set of modes is provided the Supplemental Material~\cite{SuppMat}.}
  \label{fig:2}
\end{figure}

This naturally brings the need for an alternative version of the LL. To do so,
we first use a symmetrized version of the classical wave equation in order to convert it into an equivalent Schr\"{o}dinger equation. The transformation implies a simple mapping between the two systems as we set $\psi_n(x)=\sqrt{z(x)}\varphi_n(x)$, which leads to
\begin{equation}
-\frac{1}{\sqrt{z(x)}}\nabla\cdot\left(z(x)\nabla \frac{\psi_n(x)}{\sqrt{z(x)}}\right)= E_n \psi_n(x)\,.
\label{eq:1118}
\end{equation}
Expanding the LHS of Eq.(\ref{eq:1118}), one obtains the Laplacian operator  and the equivalent potential
\begin{equation}
V_\mathrm{eq}(x)=\frac{1}{4}\left(2\frac{z''}{z}-\frac{|z'|^2}{z^2}  \right)\,,
\label{eq:7}
\end{equation}
see Supplemental Material~\cite{SuppMat} for details of the transformation and the way we compute $V_\mathrm{eq}(x)$.
Such a transformation is notably known as the Webster's transformation~\cite{webster19a,martin04a}. It preserves the energy spectrum and only affects the eigenmodes' shape, here modulated by the factor $\sqrt{z}$, see Fig.~\ref{fig:2}(a) for the first transformed mode. Hence, a localized mode modulated  by the delocalized function $\sqrt{z(x)}$ remains localized in the transformed basis. This transformation points out a limitation of the LLT, as one can construct, as we did, a random quantum potential leading to high frequency localization and consequently to an uninformative LL.

We thus adapt the LL by adding an energy shift to the operator in Eq.(\ref{eq:1118}), affecting the eigenvalues but not the eigenstates, and we compute the corresponding LL $\bar{u}_\psi(x;E_s)$
\begin{equation}
-\frac{1}{\sqrt{z(x)}}\nabla\cdot\left(z(x)\nabla \frac{\bar{u}_\psi(x)}{\sqrt{z(x)}}\right)+E_s \bar{u}_\psi(x)= 1\,.
\label{eq:799}
\end{equation}
Expressing this equation back into the original space through inverse Webster's transform, we obtain the final expression for the adapted LL $\bar{u}_\varphi(x;E_s)$:
\begin{equation}
-\nabla\cdot(z(x)\nabla \bar{u}_\varphi(x)) + E_s \bar{u}_\varphi(x) z(x)=\sqrt{z(x)} \,.
\label{eq:799}
\end{equation}
We therefore derive an adapted supremum for the original classical modes 
\begin{equation}
\bar{R}^\varphi_{n}\overset{\text{def}}{=}\frac{|\varphi_n(x)|}{||\varphi_n(x)||_{L^\infty} (E_n + E_s)\sqrt{z(x)}}\leq \bar{u}_\varphi(x)\,,
\label{eq:399}
\end{equation}
and an adapted expression for their support
\begin{equation}
\bar{S}^\varphi_{n}\overset{\text{def}}{=}\left\{ x \in \Omega : \bar{u}_\varphi(x) \geq \frac{1}{(E_n+E_s)\sqrt{z(x)}} \right\}\,.
\label{eq:8787844}
\end{equation} 
Unlike their previous versions and because of the extra $\sqrt{z(x)}$ term, Eqs.(\ref{eq:399}) and (\ref{eq:8787844}) are now based on a local energy criteria to bound a mode or to determine its support. The support for the first transformed mode $\bar{S}_1^\varphi$, computed according to Eq.(\ref{eq:8787844}) for a given $E_s$, is shown in Fig.~\ref{fig:2}(a). Given that all eigenmodes form a basis of orthogonal functions $\langle \varphi_n |\varphi_m \rangle =\delta_{nm}$, the mode $\varphi_m$ occupying the domain $\Omega_m$ is thus orthogonal to all the previous modes $\varphi_n$ for $n<m$, which previously occupied a smaller subdomain $\Omega_n \in \Omega_m$. If, at a given energy $E_m$, the support $\bar{S}^\varphi_m$ suddenly allows for the occupation of a new subdomain $\Omega_m \subset \Omega$, relatively to the previous eigenmode $\varphi_{m-1}$, it is then likely that the mode $\varphi_m$ will be localized in the newly allowed subdomain $\Omega_m \setminus \Omega_{(m-1)}$. This is indeed a more energetically favorable configuration to satisfy the mode orthogonality condition. This is illustrated in Fig.\ref{fig:2}(b,c), where we plot two high frequency localized modes with the components of the inequality in Eq.(\ref{eq:8787844}), defining their support. For the first localized mode at $n=66$, the energy level $1/(E_{66}+E_s)\sqrt{z}$ locally appears below the main peak of the LL $\bar{u}_\varphi$, thus allowing for the new subdomain $\Omega_{66} \setminus \Omega_{65}$ to participate as a support for the mode $\bar{R}^\varphi_{66}$, see red-shaded area. As the eigenvalue increases, the same process occurs for the LL's secondary peaks, see Fig.~\ref{fig:2}(c), generating other high energy Anderson localized modes. This process is perfectly captured by the proposed extension of the LLT.

We emphasize that the adapted LL $\bar{u}_\varphi$ can be computed for an arbitrary energy shift $E_s$. Yet, the choice of $E_s$ has a strong influence on the LL's final shape. For a small shift ($E_s \rightarrow 0$), the LL remains uninformative as it is akin to a global structure mode~\cite{fouque_book07a}, as in Fig.~\ref{fig:1}(h). In the limit of large shifts ($E_s \rightarrow +\infty$) the LL converges to the medium's structure ($\bar{u}_\varphi \rightarrow 1/\sqrt{z}$). In the first limit the LL does not possess any peaks or valleys, and in the second limit all peaks or valleys have the same values, which is uninformative in both cases. Only intermediary values of $E_s$ lead to a useful landscape with distinct local maxima, as shown in Fig.~\ref{fig:2}(d) where $\bar{u}_\varphi(x;E_s)$ is plotted for different values of $E_s$. 

To finally get rid of the arbitrary choice of $E_s$, we propose a method to compute a minimal support for each mode. We have seen that a mode $\varphi_n$ must be mostly located in the support $\bar{S}^\varphi_{n}$ defined by Eq.(\ref{eq:8787844}), and this, for all values of $E_s$. By contraposition, we can state that if, at a given position $x$, there exists a shift $E_s$ for which Eq.(\ref{eq:8787844}) is not satisfied, then 
$x$ must be excluded from $\bar{S}_{\varphi,n}$. So for every point of the domain $\Omega$ and at a given energy level $E_n$ of interest, one can simply check among a previously calculated set of landscape functions $\bar{u}_\varphi(x;E_s)$, if there is a value $E_s$ for which Eq.(\ref{eq:8787844}) is not satisfied. In such case this position is excluded from the support of the mode $\varphi_n$. This procedure amounts in computing the intersection of all mode supports associated with the considered shifts, and it defines the minimal support for the modes
\begin{equation}
\bar{S}^{\varphi,\mathrm{min}}_{n}=\left\{ x \in \Omega :u_\varphi^\star(x) \geq \frac{1}{E_n} \right\}\,,
\label{eq:87878}
\end{equation} 
with
\begin{equation}
u_\varphi^\star(x)=\underset{E_s}{\mathrm{max}} \left\{\left(\frac{1}{\bar{u}_\varphi (x;E_s) \sqrt{z(x)}}-E_s\right)^{-1}\right\}\,.
\label{eq:1458}
\end{equation}
In Fig.\ref{fig:2}(e) we show the measure of the mode support $\Sigma_n(E_s)=\int_\Omega\bar{S}^\varphi_{n}(E_s) dx$ for different values of shift, along with the measure of the minimal support $\Sigma_n^\mathrm{min}=\int_\Omega\bar{S}^{\varphi,\mathrm{min}}_{n} dx$. This highlights the drawback in using a single shift value $E_s$: the landscape $\bar{u}_\varphi(x;E_s)$ and support $\bar{S}^\varphi_{n}(E_s)$ only provide useful information in a certain range of energies (or mode number), close to $E_s$. Above this value, $\Sigma_n(E_s)$ converges to the whole domain $\Omega$, hence being not informative anymore. 

The function $u_\varphi^\star$ defining the minimal support and the example of the minimal support for the $66^\mathrm{th}$ mode are presented in Fig.~\ref{fig:2}(f). It clearly shows how low energy modes possess an extended support (purple-shaded areas) and how extra supports (red-shaded area) for the localized mode appear at higher energies. Equations (\ref{eq:87878}) and (\ref{eq:1458}) arise as a powerful tool to predict the position of high energy localized states as well as their minimal energy. 
\\
\\
In conclusion we have shown that, although being mathematically valid for classical waves, the LLT is not appropriate to describe high-energy Anderson localization. Such configuration can also occur for quantum waves with a specific potential, such as the one we derived from a single scalar wave parameter. The inversion of the usual localization/delocalization phenomenology greatly complicates the analysis of localized modes and the prediction of their position and shape. Nonetheless, we have introduced a modified LL, combining a non-uniform excitation of the system and a positive energy shift, which allows to recover information on localized modes. The minimal support for each mode can then be constructed by combining landscape functions computed for different energy shifts. This is of course computationally more expensive than solving a single elliptic problem but remains, in practice, an advantageous alternative to solving the original eigenvalue problem~\cite{golub_book13a}. 
We also emphasize that the acoustic wave equation with $\kappa\neq \rho$ could be tackled in a general way as a position-dependent mass Schr\"{o}dinger equation, and with a transformation involving a change of coordinates~\cite{plastino99a,pena04a}. 

\textit{Acknowledgments--}
This research was supported by the Excellence Initiative of Aix-Marseille University - A*Midex, a French "Investissements d'Avenir" program, and the Région Sud (project AndaLoca n°$2020\_03026$).

%

\pagebreak

\onecolumngrid
\begin{center}
  \textbf{\large Supplemental Material : Crossover Between Quantum and Classical Waves and High Frequency Localization Landscapes}\\[.2cm]
  David Colas,$^{1}$ Cédric Bellis,$^{1}$ Bruno Lombard$^1$ and Régis Cottereau$^{1,\dagger}$\\[.1cm]
  {\itshape ${}^1$Aix Marseille Univ, CNRS, Centrale Marseille, LMA UMR 7031, Marseille, France\\}
  ${}^\dagger$Electronic address: cottereau@lma.cnrs-mrs.fr\\
(Dated: \today)\\[1cm]
\end{center}

\twocolumngrid

\setcounter{equation}{0}
\setcounter{figure}{0}
\setcounter{table}{0}
\setcounter{page}{1}
\renewcommand{\theequation}{S\arabic{equation}}
\renewcommand{\thefigure}{S\arabic{figure}}
\renewcommand{\bibnumfmt}[1]{[S#1]}
\renewcommand{\citenumfont}[1]{S#1}

In this Supplemental Material we provide additional details on the Webster's transformation and the potential function used in the main text, and we present a 2D example showing the effect of the boundary conditions on the mode localization. Finally, we describe the content of the two Supplementary Videos. Equations and figures from the main text are here quoted with numbers, whereas those from the supplemental material are prefixed by ``S''.

\subsection{Effect of boundary conditions on 2D wave localization}

\begin{figure}[t!]
  \includegraphics[width=\linewidth]{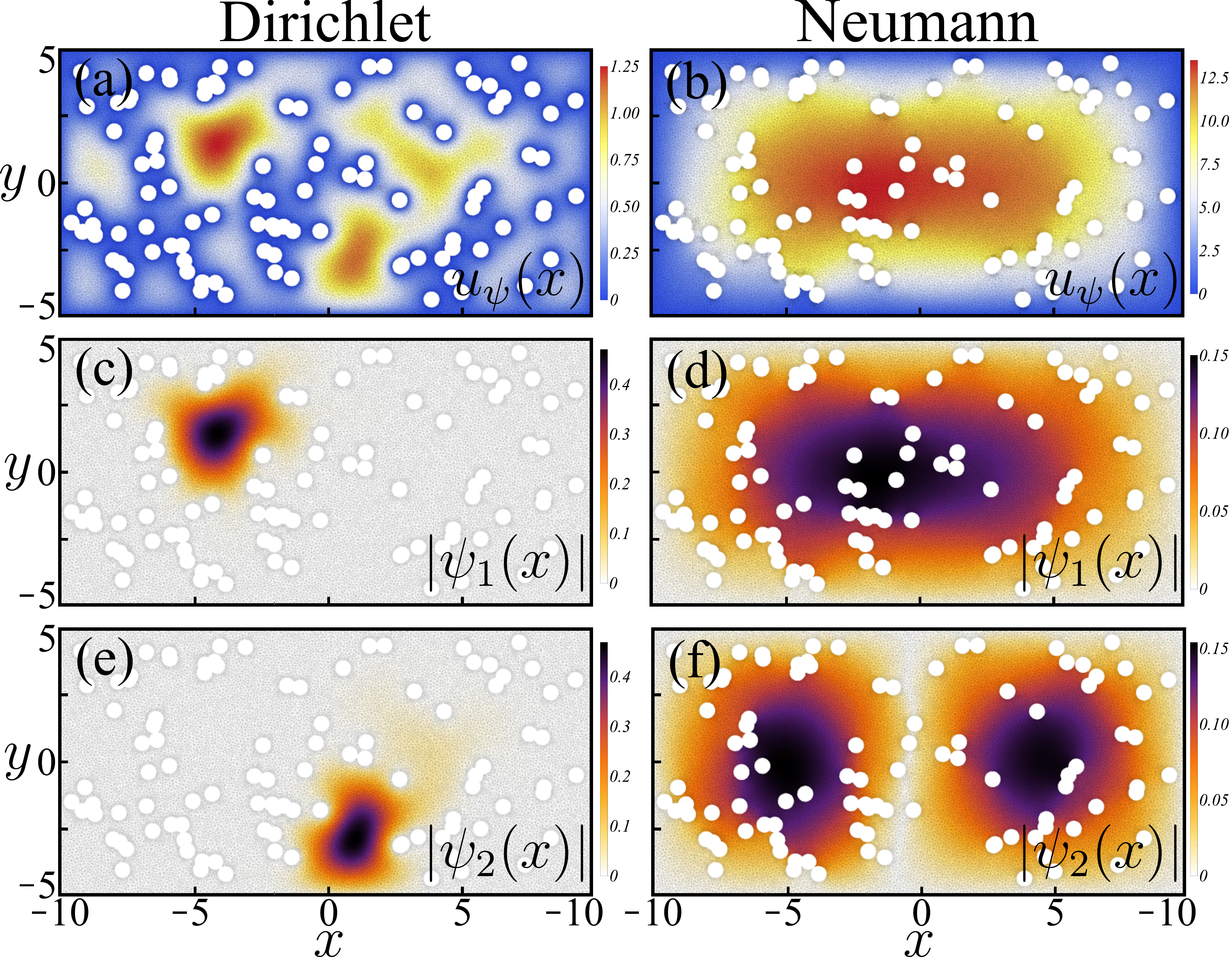}
  \caption{ 
  Effect of boundary conditions on mode localization for a 2D homogeneous surface randomly pierced with holes.
Left column corresponds to homogeneous Dirichlet BCs on all edges. Right column corresponds to homogeneous Dirichlet BCs on the edges of domain and homogeneous Neumann BCs on the edges of the holes. (a,b) Localization landscape $u_\psi$. First (c,d) and second (e,f) eigenmodes.}
  \label{fig:S1}
\end{figure}

We provide a concrete and illustrative example of the effect of the boundary conditions (BCs) on the mode localization for a homogeneous 2D surface randomly pierced with holes. We always assume homogeneous Dirichlet BCs on the edge of the 2D domain and we either consider homogeneous Dirichlet or homogeneous Neumann BCs on the edges of the holes. Using a finite-element method, we compute the first modes for the eigenvalue problem $-\Delta \psi_n = E \psi_n$ and the associated localization landscapes $-\Delta  u_\psi =1$, with the corresponding BCs. The localization landscapes along with the two first modes for each case are shown in Fig.~\ref{fig:S1}.
The case with homogeneous Dirichlet BCs everywhere is analog to the one presented in [15], and exhibits low frequency mode localization, with the main peaks of the localization landscape indicating the position and shape of the first localized eigenmodes, see panels (a,c,e). However, with homogeneous Neumann BCs on the edges of the holes, the first modes are essentially unaffected by the presence of holes and thus they are delocalized. The localization landscape is akin to the first global structure mode, so uninformative, see panels (b,d,f).

\subsection{Webster's transformation}

The Webster's transformation maps a one-parameter classical wave equation into a Schr\"{o}dinger equation (SE). Strictly speaking, the transformation maps their associated spectral problems. Let us start from the later, a time-independent SE:
\begin{equation}
(-\Delta +V(x))\psi_n(x)= E_n \psi_n(x)\,,
\label{eq:AP1}
\end{equation}
where $E_n$ is the eigenenergy of the associated quantum eigenmode $\psi_n(x)$. We set $\psi_n(x)=y(x)\varphi_n(x)$, with $\varphi_n(x)$ the classical modes and where $y(x)$ is to be determined. Substitution into Eq.~(\ref{eq:AP1}) yields
\begin{multline}
-y(x)\Delta\varphi(x)-2\nabla y(x) \cdot \nabla\varphi(x) +\\
 [-\Delta +V(x) y(x)]\varphi(x) = E \varphi(x) \,.
\label{eq:AP2}
\end{multline}
To get rid of the linear term in $\varphi$, we define $y(x)$ as a non-trivial solution of the homogeneous differential equation 
\begin{equation}
[-\Delta +V(x) y(x)]\varphi(x)=0 \,.
\label{eq:AP3}
\end{equation}
This lets us with the equation
\begin{equation}
\Delta \varphi(x) +\nabla \ln(y^2(x))\cdot\nabla\varphi(x)=-E\varphi(x)\,.
\label{eq:AP4}
\end{equation}
Now, setting $z(x)=y^2(x)$ in the above equation, we get
\begin{equation}
z(x) \Delta \varphi(x) +\nabla z(x)\cdot \nabla \varphi(x)= -E z(x) \varphi(x)\,,
\label{eq:AP5}
\end{equation}
which can be recast as
\begin{equation}
-\frac{1}{z(x)}\nabla\cdot(z(x)\nabla \varphi(x))=E\varphi(x)\,,
\label{eq:AP6}
\end{equation}
that is a one-parameter classical wave equation, and also called Webster's horn equation.

Also, given Eq.~(\ref{eq:AP3}) and that $z(x)=y(x)^2$, we can derive a unique expression for the equivalent Schr\"{o}dinger potential which reads
\begin{equation}
V_\mathrm{eq}(x)=\frac{1}{4}\left(2\frac{z''}{z}-\frac{|z'|^2}{z^2}  \right)\,.
\label{eq:AP7}
\end{equation}
Reciprocally, if one wants, from a given potential $V(x)$, to determine the corresponding classical parameter $z(x)$, one needs to solve Eq.~(\ref{eq:AP3}) for $z(x)$, which requires two integration constants.

\subsection{Acoustic structure and associated potential}
\begin{figure}[t!]
  \includegraphics[width=\linewidth]{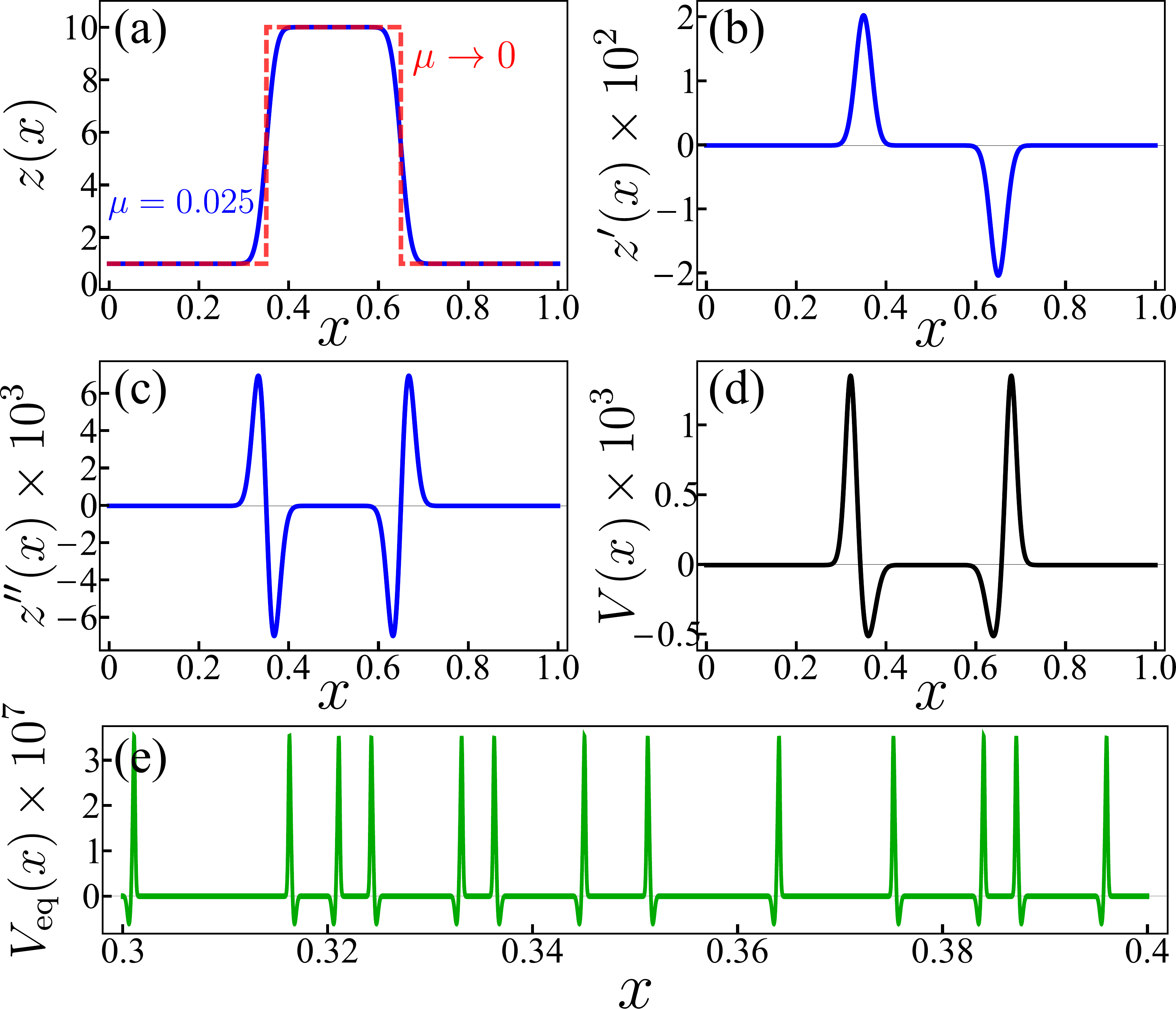}
  \caption{(a) Example of a basis function $z(x)$, centered at $x_k=0.5$, with $z_\mathrm{max}=10$, $z_\mathrm{min}=1$ and a width $\sigma=0.3$. With $\mu=0.025$ edges are smoothed (blue line), and the original rectangle function is recovered with $\mu\rightarrow 0$ (dashed-red line). (b) First derivative of $z(x)$, that is a sum of two Gaussian functions of opposite sign. (c) Second derivative of $z(x)$. (d) Quantum potential obtained from $z(x)$, see Eq.(\ref{eq:AP7}), possessing both a positive and negative part. (e) Section of the effective potential computed with parameters from the main text.}
  \label{fig:supp1}
\end{figure}

The Bernoulli-type random parameters from the main text essentially consist in a succession of rectangle functions with different widths. Practically, to numerically construct our classical 1D random structure $z(x)$, we choose a smoothed approximation of rectangular functions based on \textit{error} functions. At a site $k$, we thus have
\begin{multline}
z_k(x)=\frac{z_\mathrm{max}-z_\mathrm{min}}{4}\Bigg[\mathrm{erf}\left(\frac{x+\sigma_k/2 -x_k}{\mu}  +1\right)\\
\mathrm{erf}\left(\frac{-x+\sigma_k/2 +x_k}{\mu} +1 \right)\Bigg]\,,
\label{eq:ST2}
\end{multline}
shaping a smooth rectangle function of height $z_\mathrm{max}$, centered at a position $x_k$ and with width $\sigma_k$. The parameter $\mu$ controls the smoothness of the rectangle's walls, assuming $0<\mu\ll \sigma$. In the limit of $\mu\rightarrow 0$, $z_k(x)$ defines a straight rectangle function. This choice is particularly convenient for the derivation and the numerical computation of the Webster's potential $V_\mathrm{eq}(x)$, see Eq.~(\ref{eq:AP7}), since it essentially involves first and second derivatives of $z(x)$. They end up being Gaussian and first derivative of Gaussian functions, instead of Dirac delta functions and their derivatives if straight rectangle functions are chosen. The rectangle $(\mu\rightarrow 0)$ and smoothed rectangle functions $(\mu > 0)$ are plotted in Fig.~\ref{fig:supp1}(a), along with the first (b) and second derivative (c), for the smoothed case. The potential function from Eq.~(\ref{eq:AP7}) is shown in Fig.~\ref{fig:supp1}(d). A section of the equivalent potential computed from the main text's parameter $z(x)$ is also shown in Fig.~\ref{fig:supp1}(e). Performing the Webster's transform here leads to an equivalent potential which resembles a complicated ``'Gaussian'' comb. 

\subsection{Supplementary Videos}

Two videos are provided. Supplementary video S1 corresponds to an animated version of Fig.1(b,d,f,h) for the first 100 eigenmodes of the quantum (top) and classical (bottom) systems. Normalized modes $R_{\psi,n}$ and $R_{\varphi,n}$ with their respected localization landscapes and supports are shown on the left. Eigenmodes simply normalized to 1 are shown on the right. Supplementary video S2 corresponds to an animated version of Fig.2(a,b,c,f), also for the first 100 eigenmodes. The first two rows show the adapted landscape $\bar{u}_\varphi(x;E_s)$ and the mode support defined by the local condition defined in Eq.(12), with a focus on specific areas. The bottom row shows the function $u^\star(x)$ defining the minimum support and its evolution as function of $1/E_n$ decreases.
\end{document}